\documentclass[]{spie}  

\usepackage{amsmath,amsfonts,amssymb}
\usepackage{graphicx}
\usepackage[colorlinks=true, allcolors=blue]{hyperref}

\usepackage{float}
\usepackage{array}
\usepackage{booktabs}
\usepackage{adjustbox}
\usepackage{multirow}

\newcolumntype{P}[1]{>{\centering\arraybackslash}p{#1}}
\newcolumntype{M}[1]{>{\centering\arraybackslash}m{#1}}

\title{Medical Knowledge-Guided Deep Curriculum Learning for Elbow Fracture Diagnosis from X-Ray Images}

\author[a]{Jun Luo}
\author[b]{Gene Kitamura}
\author[b]{Emine Doganay}
\author[b]{Dooman Arefan}
\author[a, b, c, d]{Shandong Wu}
\affil[a]{Intelligent Systems Program, University of Pittsburgh, 4200 Fifth Avenue, Pittsburgh, PA, USA, 15213}
\affil[b]{Dept. of Radiology, University of Pittsburgh, 4200 Fifth Avenue, Pittsburgh, PA, USA, 15213}
\affil[c]{Dept. of Biomedical Informatics, University of Pittsburgh, 4200 Fifth Avenue, Pittsburgh, PA, USA, 15213}
\affil[d]{Dept. of Bioengineering, University of Pittsburgh, 4200 Fifth Avenue, Pittsburgh, PA, USA, 15213}

\authorinfo{Correspondence to S. Wu. E-mail: wus3@upmc.edu, Telephone: 1 412 641 2567}

\begin{document} 
\maketitle

\begin{abstract}
Elbow fractures are one of the most common fracture types. Diagnoses on elbow fractures often need the help of radiographic imaging to be read and analyzed by a specialized radiologist with years of training. Thanks to the recent advances of deep learning, a model that can classify and detect different types of bone fractures needs only hours of training and has shown promising results. However, most existing deep learning models are purely data-driven, lacking incorporation of known domain knowledge from human experts. In this work, we propose a novel deep learning method to diagnose elbow fracture from elbow X-ray images by integrating domain-specific medical knowledge into a curriculum learning framework. In our method, the training data are permutated by sampling without replacement at the beginning of each training epoch. The sampling probability of each training sample is guided by a scoring criterion constructed based on clinically known knowledge from human experts, where the scoring indicates the diagnosis difficultness of different elbow fracture subtypes. We also propose an algorithm that updates the sampling probabilities at each epoch, which is applicable to other sampling-based curriculum learning frameworks. We design an experiment with 1865 elbow X-ray images for a fracture/normal binary classification task and compare our proposed method to a baseline method and a previous method using multiple metrics. Our results show that the proposed method achieves the highest classification performance. Also, our proposed probability update algorithm boosts the performance of the previous method.
\end{abstract}


\section{INTRODUCTION}
\label{sec:intro}  

Elbow fracture is one of the fracture types that happens most frequently among people across all ages. It is crucial for patients with elbow fracture to have timely diagnosis and treatment since the fracture could cause neurovascular damage\cite{saeed2017elbow}. Radiographic imaging, such as X-ray, is one of the most popular image modalities to help provide physicians with prompt assessment of the patient since it can help visualize the interior of the patient’s elbow in a timely manner. Traditionally, it takes a physician years of training in order to diagnose elbow fractures from radiographic images. In recent years, thanks to the thrive of deep learning, a model that can classify and detect of different types of bone fractures can be trained within hours from annotated images to help diagnose elbow fractures\cite{rayan2019binomial,england2018detection,tanzi2020x,jimenez2019medical}. However, most existing methods purely rely on the data-driven learning, lacking leveraging known medical knowledge that physicians acquired after years of training.

In this work, we propose a novel deep learning framework that incorporates medical knowledge from domain experts (i.e., radiologists) through a curriculum learning\cite{bengio2009curriculum} method to augment the data-driven learning on a binary (fracture vs. normal) classification of the elbow fractures. Curriculum learning was first introduced to allow the machine to mimic human learning and train a model by feeding examples in an easy-to-hard order, opposite to the generally used random order\cite{bengio2009curriculum}. Since then, curriculum learning has been used in many areas such as image classification\cite{jimenez2019medical,gong2016multi,lotter2017multi,wang2019dynamic,wei2021learn}, object detection\cite{zhang2019leveraging,zhang2017bridging}, semantic segmentation\cite{zhang2017curriculum,kervadec2019curriculum}, self/semi supervised learning\cite{gong2016multi,zhang2019leveraging,zhang2017bridging,murali2018cassl,tang2018attention,kervadec2019curriculum}, multi-task learning\cite{sarafianos2017curriculum,sarafianos2018curriculum,dong2017multi,wang2018mancs}, multi-modal learning\cite{gong2016multi,lotter2017multi,gong2017exploring}, etc\cite{jiang2015self,matiisen2019teacher,weinshall2018curriculum}. However, few works take advantage of the outside knowledge from human experts, which could be particularly helpful in the field of medical imaging since human experts are trained for years.

\begin{figure} [t]
   \begin{center}
   \begin{tabular}{c} 
   \includegraphics[width=0.97\textwidth]{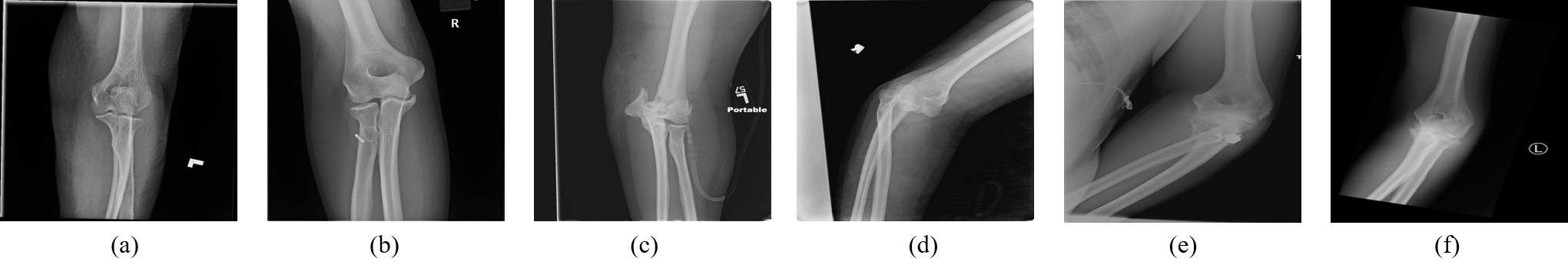}
   \end{tabular}
   \end{center}
   \caption[]
   {Six Subtypes of elbow fractures: (a) Ulnar fracture; (b) Radial fracture; (c) Humeral fracture; (d) Dislocation; (e) Complex fracture/multi-type fracture; (f) Coronoid process fracture.\label{fig:expfrac}}
\end{figure} 

In our proposed framework, we design a scoring criterion, constructed based on clinically known knowledge from human experts, that indicates the diagnosis difficultness of different elbow fracture subtypes. The curriculum is then created through permutations of the training data by sampling without replacement at the beginning of each epoch and the scoring criterion is used to guide the sampling probability of each training sample. We also propose an algorithm that updates the sampling probabilities at each epoch, and this update algorithm is applicable to other sampling-based curriculum learning frameworks. 

To evaluate our method, we design an experiment with an elbow X-ray dataset of 1865 patients and compare the effects with a baseline method and a previous method. Results show that our method outperforms the compared methods, and our proposed probability update algorithm boosts the performance of the previous method.

The remainder of this paper is organized as follows: Section \ref{sec:methods} provides the details of the study cohort and the proposed method; Results of the experiment are shown in Section \ref{sec:results}; Section \ref{sec:discussion} discusses our findings and limitations and concludes the study.

\section{METHODS}
\label{sec:methods}

\subsection{Study cohort and dataset}
\begin{table}[H]
    \caption{Number of images of the normal cases and six subtypes of the elbow fractures.} \label{table:dataset}
    \centering
    \begin{tabular}{M{1.0cm}M{1.5cm}M{0.8cm}M{0.8cm}M{0.8cm}M{0.8cm}M{0.8cm}M{0.8cm}M{2.6cm}M{0.9cm}}
    \toprule
    Type & (\textit{normal}) & (\textit{a}) & (\textit{b}) & (\textit{c}) & (\textit{d}) & (\textit{e}) & (\textit{f}) & Total of (\textit{a})-(\textit{f}) \newline{(\textit{fracture})} & Total\\ \midrule
    Train & 800 & 88 & 340 & 84 & 11 & 42 & 27 & 592 & 1,392\\
    Test & 400 & 10 & 44 & 9 & 2 & 4 & 4 & 73 & 473\\
    Total & 1,200 & 98 & 384 & 93 & 13 & 46 & 31 & 665 & 1,865\\
    \bottomrule
    \end{tabular}
\end{table}
In this study, we use a cohort of 1,865 frontal view elbow radiographic images of patients with elbow trauma, where 665 are fracture positive cases and 1,200 are non-fracture (i.e., normal) cases. The elbow radiographic images were retrospectively collected at our institution and reviewed by expert radiologists. Among the 665 fracture cases, each case is annotated with a specific subtype of fracture out of a total of six different subtypes, as demonstrated in Figure \ref{fig:expfrac}. We focus on a binary classification task by grouping all six different subtypes of fractures as class fracture and assigning the non-fracture cases as class normal. We split the dataset into training set and test set while preserving a consistent percentage of each fracture subtype in the training and test set. Table \ref{table:dataset} shows the distribution of our image dataset. During testing of the models, we evaluate effects on 100 randomly selected normal cases from the test set and all 73 fracture cases from the test set that serve as a more balanced test between the binary classes, and we repeat the test 5 times, each time selecting different set of normal cases from the test set, to increase the robustness of the evaluation and report the average experiment results.  We hereafter use the following notations to denote the six subtypes of elbow fracture: (a) for ulnar fracture; (b) for radial fracture; (c) for humeral fracture; (d) for dislocation; (e) for complex fracture/multi-type fracture; (f) for coronoid process fracture.

\subsection{Knowledge-based curriculum}
The idea of curriculum learning is first brought up to let the machine imitate human learning by first learning with “easier” samples, and then transit to learning from “harder” samples in the entire training set. In the process of constructing a curriculum, we need to design a quantifiable criterion that reflects how hard it is to classify a certain subtype of elbow fracture in clinical practice, and then utilize the criterion to schedule the order of feeding the samples into the model for training. Here, we use radiologists’ accumulated clinical experience on the difficultness of diagnosis of the elbow subtype fractures as a medical knowledge to formalize the criterion. As shown in Table \ref{table:diff}, a board-certified experienced musculoskeletal radiologist (G.K.; $>10$ years of experience) was asked to score the general level of difficulty of the elbow fracture diagnosis from the frontal radiographic view images of the elbow, with the scores ranging from 1 (hardest) to 100 (easiest).

\begin{table}[H]
    \caption{Number of images of the normal cases and six subtypes of the elbow fractures.} \label{table:diff}
    \centering
    \begin{tabular}{M{1.0cm}M{2.0cm}M{1.7cm}M{1.7cm}M{1.7cm}M{1.7cm}M{1.7cm}M{1.7cm}}
    \toprule
    & (\textit{normal}) & (\textit{a}) & (\textit{b}) & (\textit{c}) & (\textit{d}) & (\textit{e}) & (\textit{f})\\ \midrule
    Score & 30 & 30 & 30 & 70 & 40 & 90 & 10\\
    \bottomrule
    \end{tabular}
\end{table}

In the training process, consider a triplet $\left(x_i,y_i,f_i\right)\in\{(x_j,y_j,f_j)\}_{j=1}^N$ in the training set, where $x_i$ is the $i$-th image in the training set $\mathcal{X}$ that contains $N$ images, $y_i\in\mathcal{Y}=\{0,1\}$ is its label, and $f_i\in\mathcal{F}=\{\left(normal\right),\left(a\right),\left(b\right),\left(c\right),\left(d\right),\left(e\right),\left(f\right)\}$ is its fine-grained label, indicating which subtype of the elbow fracture is, if it is not $\left(normal\right)$. As mentioned above, we focus on a fracture vs. normal binary classification task and $y_i=0$ if and only if $f_i=\left(normal\right)$.

We design our curriculum with a sampling-without-replacement strategy to permute and reorder the training set at the beginning of each training epoch. For sampling, each $x_i$ is given a probability, $p_i$, whose value is updated at the beginning of each training epoch, with the initial value $p_{i,\left(1\right)}$ assigned at the first epoch. Note that the probability update method mentioned below can be utilized by any curriculums learning framework that is based on permuting the training set by the strategy of sampling-without-replacement.

\paragraph{Initialization of $p_i$}
The $p_{i,\left(1\right)}$ for all images are computed from the hardness scores as provided in Table \ref{table:diff}. Our curriculum addresses the importance of learning from easier cases, or, here, cases with higher scores by defining the initial probability $p_{i,\left(1\right)}$ for image $x_i$ according to Equation (\ref{eq:diff2prob}), where $s_{f_k}$ is the score for image $x_k$ in terms of its fine-grained class label $f_k$, and N is the total number of images in the training set.

\begin{equation}
    \label{eq:diff2prob}
    p_{i,\left(1\right)}=\frac{s_{f_i}}{\sum_{j=1}^{N}s_{f_j}}
\end{equation}

\paragraph{Update of $p_i$}
At the beginning of each epoch after the first, the $p_i$’s are updated by Equations (\ref{eq:lamb}) and (\ref{eq:p_updtae}): 
\begin{equation}
    \label{eq:lamb}
    \lambda_i=\sqrt[L]{\frac{p_{i,\left(final\right)}}{p_{i,\left(1\right)}}}=\sqrt[L]{\frac{1/N}{p_{i,\left(1\right)}}} 
\end{equation}

\begin{equation}
    \label{eq:p_updtae}
    p_{i,\left(e\right)}=\left\{\begin{matrix}\begin{matrix}p_{i,\left(e-1\right)}\cdot\lambda_{i,}\\1/N,\\\end{matrix}&\begin{matrix}2\le e\le L\\L<e\le E\\\end{matrix}\\\end{matrix}\right. ,
\end{equation}
where $p_{i,\left(e\right)}$ is the probability for image $x_i$ after the update at epoch $e$, and $\lambda_i$ is a scalar for $x_i$ that ensures $p_i$ for every image to exponentially and smoothly transit to $p_{i,\left(final\right)}=1/N$. $E$ denotes the total number of epochs, and $L$ is the number of the epoch after which $p_i$ equals to $p_{i,\left(final\right)}$ for the rest of the training process. We treat $L$ as a hyperparameter. In this way, the model can gradually reduce the emphases we initially put on the easier cases, and then reach to a stage of random selection of samples in the entire training set after epoch $L$.

\section{RESULTS}
\label{sec:results}
\begin{figure} [t]
   \begin{center}
   \begin{tabular}{c} 
   \includegraphics[width=0.8\textwidth]{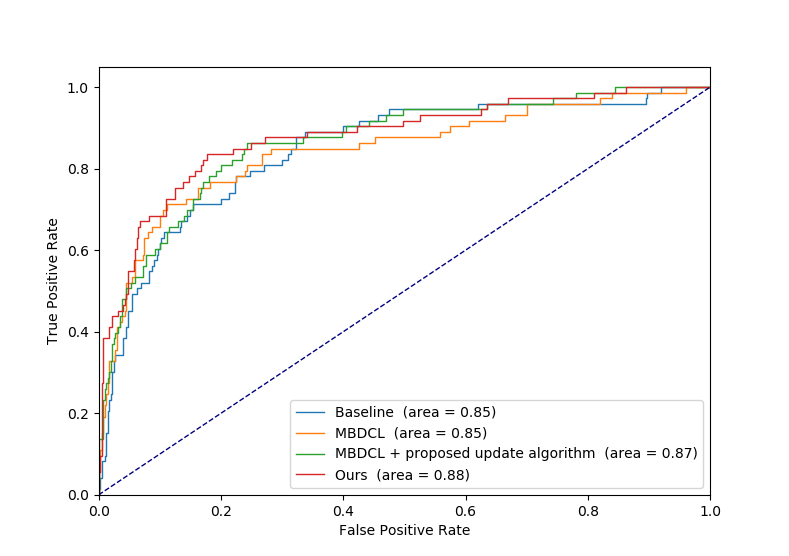}
   \end{tabular}
   \end{center}
   \caption[]
   {ROC curves of different methods for the entire test set.\label{fig:roc}}
\end{figure}

We evaluate the performance of our curriculum on a total of 1,865 X-ray images (each image from a different patient). For each hyperparameter setting, we train the model 5 times to reduce potential bias from the randomness of sampling and training initialization. Results reported below are means and standard deviations across the 5 runs. All experiments are implemented in PyTorch\cite{paszke2019pytorch} framework on an NVIDIA TESLA V100 GPU from Pittsburgh Supercomputing Center, with the same effort for hyperparameter tuning and data augmentation. 
We compare our method with two other methods: 1) a baseline method, which is the most commonly used method of random shuffling the order of the images in the training set; and 2) a previously reported curriculum learning method (here referred as MBDCL\cite{jimenez2019medical}) that uses a different way to permute the training set. In addition, in order to evaluate the effects under the opposite strategy of our curriculum, we also compare our method to an anti-curriculum setting where we reverse the difficultness scoring (listed in Table \ref{table:diff}). We reverse the difficultness scoring by re-assigning the scores as 100 minus the original scores, indicating difficultness information that is the opposite to the knowledge provided by the radiologist. Furthermore, we evaluate the effects of plugging our probability update algorithm to the previous method (MBDCL + proposed update algorithm) and the anti-curriculum method (Anti-curriculum + proposed update algorithm). To ensure a fair comparison, all the methods use the same backbone of VGG16\cite{simonyan2014very} network architecture. The evaluation metrics include accuracy, area under the ROC curve (AUC), Average Precision, and F1-score. Our experiment results are shown in Table \ref{table:results}. The ROC curves for the entire test set are shown in Figure \ref{fig:roc}.

\begin{table}[H]
    \caption{Performance comparison of different methods.} \label{table:results}
    \centering
    \begin{tabular}{wl{7.0cm}|M{2.0cm}M{2.0cm}M{2.0cm}M{2.0cm}}
    \toprule
    & \multicolumn{4}{c}{\small{Average on 5 Different Test Subsets (100 normal + 73 fracture)}} \\
    & Accuracy & AUC & Average Precision & F1 score\\ \midrule
    Baseline & $0.776 \pm 0.026$ & $0.834 \pm 0.025$ & $0.788 \pm 0.043$ & $0.716 \pm 0.032$ \\
    MBDCL\cite{jimenez2019medical} & $0.797 \pm 0.025$ & $0.865 \pm 0.019$ & $0.831 \pm 0.029$ & $0.763 \pm 0.022$ \\
    MBDCL\cite{jimenez2019medical} + proposed update algorithm & $0.806 \pm 0.027$ & $0.878 \pm 0.022$ & $0.847 \pm 0.035$ & $0.762 \pm 0.029$ \\
    \textbf{Ours} & $\pmb{0.809 \pm 0.023}$ & $\pmb{0.882 \pm 0.020}$ & $\pmb{0.852 \pm 0.036}$ & $\pmb{0.778 \pm 0.024}$ \\\midrule
    Anti-curriculum & $0.765 \pm 0.032$ & $0.863 \pm 0.020$ & $0.821 \pm 0.036$ & $0.757 \pm 0.025$ \\
    Anti-curriculum + proposed update algorithm & $0.803 \pm 0.036$ & $0.871 \pm 0.025$ & $0.838 \pm 0.045$ & $0.773 \pm 0.031$ \\
    \bottomrule
    \end{tabular}
\end{table}

\section{DISCUSSION AND CONCLUSION}
\label{sec:discussion}

In this work, we propose a novel medical knowledge-guided deep curriculum learning method for elbow fracture diagnoses from X-ray images. The knowledge is incorporated into the modeling process in the form of a pre-defined quantitative scoring criterion on the difficultness of classifying the elbow fracture subtypes. Various medical knowledge exists or can be accumulated/acquired from clinical experience. It will be a more effective way to augment the pure data-driven deep learning by leveraging the medical knowledge to the models, and to further improve the performance of deep learning. In this study, we formulate a technical framework to define and incorporate radiologists’ diagnosis knowledge and we show in the results the following findings: first, our method outperforms all other compared methods (accuracy=0.81, AUC=0.88); second, the anti-curriculum settings demonstrate inferior results as expected, which reflects the positive impact of our original curriculum; in addition, we also show that the proposed probability update algorithm can further enhance other curriculum learning methods.

Despite the promising results, our work has some limitations. In our study, we are currently using a cohort of around 1900 images collected at our institution, where only around a third of which are fracture cases. A more balanced dataset and potentially multi-center larger cohorts would help further evaluate our method especially the robustness and generalizability.

There also exists some interesting topics for future investigation. Patients with elbow trauma are often suggested to take radiography from multiple perspectives (frontal and lateral views). Although plugging our method directly into a multiview deep neural network structure is feasible, more informative domain knowledge regarding different views could further be involved in the curriculum. We believe that the patterns and features learned by the model from different individual views could complement one another and possibly further improve the performance. In addition, based on different forms of clinical knowledge, customized schemes of combining multiple sources of knowledge are also worth investigation in future work.

In summary, in this work, we propose a novel strategy for incorporating medical knowledge to guide and enhance data-driven deep learning for medical applications especially for elbow fracture diagnosis. Our method demonstrates a new mechanism of defining and incorporating existing rich clinical experience and knowledge into artificial intelligence (AI) tools for clinical applications. We believe knowledge-guided AI represents a direction of the future research for AI in medical applications.

\acknowledgments 
This work was supported by National Institutes of Health (NIH)/National Cancer Institute (NCI) R01 grants (\#1R01CA193603, \#3R01CA193603-03S1, and \#1R01CA218405), and an Amazon AWS Machine Learning Research Award. This work used the Extreme Science and Engineering Discovery Environment (XSEDE), which is supported by National Science Foundation (NSF) grant number ACI-1548562. Specifically, it used the Bridges-2 system, which is supported by NSF award number ACI-1445606, at the Pittsburgh Supercomputing Center (PSC).

\bibliography{report} 
\bibliographystyle{spiebib} 

\end{document}